\documentclass{iau} 
\usepackage{graphicx, times, amsmath, epsfig, amssymb}
\usepackage{verbatim}

\newcommand{\avenf}{\bar{x}_{\rm HI}}

\newcommand{\Msun}{M_\odot}

\newcommand\lsim{\mathrel{\rlap{\lower4pt\hbox{\hskip1pt$\sim$}}
        \raise1pt\hbox{$<$}}}
\newcommand\gsim{\mathrel{\rlap{\lower4pt\hbox{\hskip1pt$\sim$}}
        \raise1pt\hbox{$>$}}}
\def\myputfigure#1#2#3#4#5%
{\vskip#5pt\makebox[0pt]{\hskip#2in
\includegraphics[width=#3\textwidth]{#1}}\vskip#4pt\hfill}

\newcommand{\apj}{ApJ}

\newcommand{\aap}{A\&A}
\newcommand{\aj}{AJ}
\newcommand{\mnras}{MNRAS}

\newcommand{\physrep}{Physics Reports}
\newcommand{\prd}{PRD}

\newcommand{\nat}{Nature}

\newcommand{\pasa}{PASA}

\title[Reionization and Cosmic Dawn Modeling] 
{Reionization and Cosmic Dawn:\\theory and simulations}

\author[Andrei Mesinger]
{Andrei Mesinger$^1$}

\affiliation{$^1$Scuola Normale Superiore, \\Piazza dei Cavalieri, 7\\  56126 Pisa, Italy\\
  email: {\tt andrei.mesinger@sns.it}}

\pubyear{2018}
\volume{333}  
\jname{Peering towards Cosmic Dawn}
\editors{Vibor Jeli\'c \& Thijs van der Hulst, eds.}
\begin{document}

\maketitle

\begin{abstract}
  We highlight recent progress in the sophistication and diversification of cosmic dawn and reionization simulations.  The application of these modeling tools to current observations has allowed us narrow down the timing of reionization, which we now know to within $\Delta z \sim 1$ for the bulk of reionization.
   The strongest constraints come from the optical depth to the CMB measured with the {\it Planck} satellite and the first detection of ongoing reionization from the spectra of the $z=7.1$ QSOs ULASJ1120+0641.  However, we still know virtually nothing about the astrophysical sources during the first billion years.  The revolution in our understanding will be led by upcoming interferometric observations of the cosmic 21-cm signal.  The properties of the sources and sinks of UV and X-ray photons are encoded in the 3D  patterns of the signal.  The development of Bayesian parameter recovery techniques, which tap into the wealth of the 21-cm signal, will soon usher in an era of precision astrophysical cosmology.
\end{abstract}

\firstsection 
\section{Introduction}

The first billion years of our Universe witnessed the birth of the first stars and galaxies.  These galactic ancestors were likely much smaller than present-day galaxies, yet they could have hosted stars and stellar black holes far more massive than typically found today.
The light emerging from these galaxies started spreading through our dark and cold Universe, heralding the {\it Cosmic Dawn} (CD).  This radiation influenced other nascent galaxies, as well as heated and ionized the pervasive intergalactic medium (IGM).  This process culminated in the final major phase change of our Universe, the {\it Epoch of Reionization} (EoR), with more than 99.99\% of the atoms in the IGM becoming ionized.  The CD and EoR contain the answers to some fundamental questions: {\it When did the first generations of galaxies form?  What were their properties?  How did they interact with each other?  What is the structure of the IGM?  What is the thermal and ionization history of the baryons?}

Answering these questions is very challenging.  Most of the first galaxies are likely far too faint to be directly observed with upcoming instruments, and their properties must be inferred indirectly.  Interpreting the current scant observations remains controversial.
The problem is two-fold.  Firstly, we are faced with a huge range of relevant scales (i.e. dynamic range).  Ultimately small-scale physics governing the birth and death of stars is responsible for driving large-scale radiation fields.  {\it The EoR/CD is inhomogeneous on scales of hundreds of Mpc}; correctly interpreting current and upcoming observations requires us to capture these inhomogeneities.  Secondly, we know very little about galaxy formation in the early Universe.  The complexities of the relevant physics combined with the lack of detailed observations leaves us with {\it an enormous parameter space of astrophysical uncertainties}.

As a result, recent years have seen a shift towards diversification of cosmological simulations of the EoR/CD.  These can sacrifice resolution and physical complexity for speed and scale, allowing us to adapt to the exigencies of each particular observation (\S \ref{sec:toolkit}).  This diverse simulation toolkit allowed us to robustly interpret the latest observations of the cosmic microwave background (CMB) as well as high-$z$ Lyman alpha emitting galaxies and quasars.  Consensus is emerging on the timing of reionization (\S \ref{sec:timing}).  However, we still know almost nothing about the details of this process and the galaxies and/or AGN thought to be driving it.  Luckily, the next few years will see a revolution in CD/EoR studies, enabled by the interferometry of the 21-cm line of neutral hydrogen (\S \ref{sec:21cm}).

\section{The toolkit for cosmological simulations}
\label{sec:toolkit}

\begin{figure*}
{
\includegraphics[width=\textwidth]{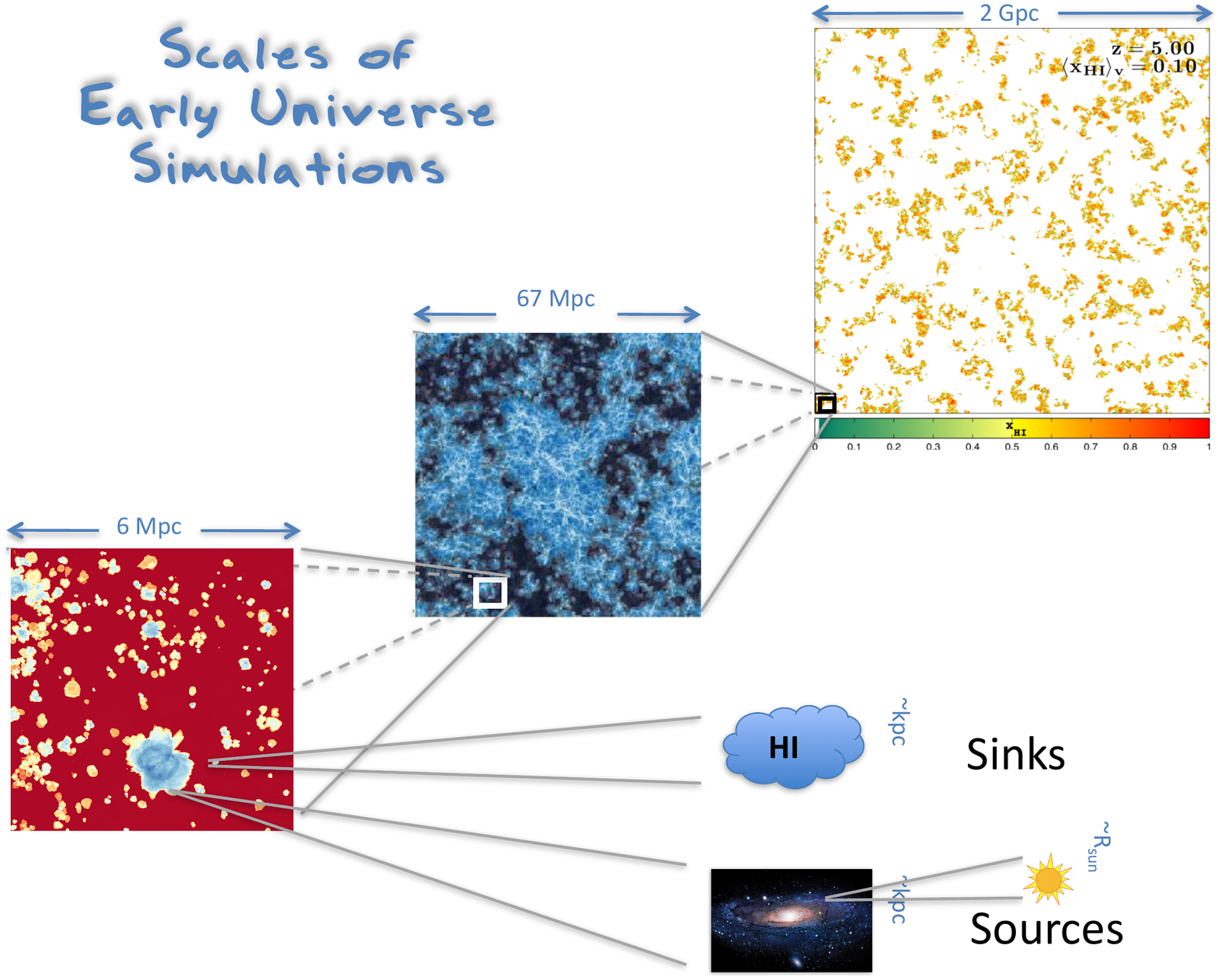}
}
\caption{
Illustration of the various scales and corresponding cosmological simulation tools.  Stars and gas clumps ultimately drive radiation fields which are inhomogeneous on cosmological scales (hundreds of Mpc).  From bottom left to top right, we show slices through ionization fields computed with hydrodynamic radiative transfer simulations (\cite{Xu16}), N-body with post-processed radiative transfer (\cite{Dixon15}), and semi-numerical simulations (\cite{Mesinger10}).  Note that the simulations are independent; zoom-ins are only to give a sense of relative scale.
}
\label{fig:scales}
\vspace{-0.5\baselineskip}
\end{figure*}

The enormous computational requirements of the EoR/CD has led to a diversification of cosmological simulation tools, as mentioned above and illustrated in Fig. \ref{fig:scales} (see also the review in \cite{TG11}).    These can vary in the treatment of both: (i) the source/sink fields (e.g. galaxies and recombining clumps), and (ii) the radiate transfer (RT).

The most realistic treatment of (i) uses coupled N-body and hydrodynamic codes, while the most realistic treatment of (ii) uses ray-tracing approaches.  Such simulations are usually limited to scales of $\sim$ 1 -- 10 Mpc, if they wish to resolve the bulk of the early galaxy population (e.g. \cite{Xu16}).  These physics-rich simulations are invaluable in resolving the very first, molecular-cooled galaxies (residing in halos with masses of $\sim10^6$--$10^8 \Msun$), as well as studying processes such as radiative feedback, metal pollution, stochastic star formation and recombining clumps in the IGM.  However, cosmological simulations still cannot resolve the sub-structure of stellar environments, turbulence and the early phases of SNe explosions, which means that results still depend on sub-grid prescriptions, albeit with fewer and more physically-motivated ``tuning knobs'' than the approaches discussed below.  Moreover, these simulations are difficult to calibrate to observations, as they usually cannot be run to moderate redshifts nor do they capture the majority of the bright galaxies we can observe.

More approximate treatments of either (i) or (ii) can extend EoR/CD simulations to $\sim$ 100 Mpc scales (e.g. \cite{TC07, Iliev14, Baek10, Ocvirk16, Kakiichi17}).  These begin to approach the scales needed to statistically sample the expected distribution of ionized regions during the EoR.  This is achieved generally by not resolving the ISM of galaxies and/or removing hydrodynamics completely.  As a result, galaxy properties such as the star formation rates and ionizing luminosities must be assigned to DM halos as input, reducing their predictive ability.  Nevertheless, given several of these input prescriptions, these simulations can be used to distinguish between the resulting ionization fields and associated observables.
Monte Carlo based RT methods allow us to achieve large scales while keeping medium resolution hydrodynamics, by reducing either the source population to only rare host dark matter or the number of emitted photon packets per galaxy (e.g. \cite{Baek10, Partl11}).  Moment based RT methods (e.g. the Optically Thin Variable Eddington Tensor approach of \cite{GA01}) achieve comparable scales with more sources, but at the cost of increased diffusion of the radiation fields.  However most RT approaches show reasonable agreement (\cite{Iliev06}).

Recent years have witnessed the advent of even more approximate, so-called ``semi-numerical'' schemes (e.g. \cite{Zahn07, MF07, GW08, CHR09}).  These replace RT of ionizing photons with an excursion-set approach (\cite{FHZ04}), in which the cumulative number of ionizing photons produced are compared against the number of neutral atoms in regions of decreasing spherical scales.  Photons with longer mean free paths, like X-rays and soft UV, can be tracked by integrating back along the light-cone (\cite{Santos10, MFC11, Fialkov13}).  Many of these schemes replace even $N$-body codes, computing the source fields directly from the density field with excursion-set halo mass functions (\cite{Bond91, LC93}).  This enables virtually unlimited dynamic range, allowing EoR/CD simulations to approximately match the field of view of upcoming 21-cm interferometers (\cite{MGS16}).  Because of their speed, semi-numerical codes enable rapid exploration of the large parameter space of astrophysical uncertainties.  However, the radiation fields become inaccurate on small scales ($\lsim$ Mpc; \cite{Zahn11}), and as with all simulations which do not resolve galactic star-formation, the inputted source properties are parametric.

Finally, ``hybrid'' techniques have even been developed, which combine some of the advantages and disadvantages of the above.  For example, the large-scale patchiness of reionization captured by semi-numerical codes can be combined (in a statistical fashion) with small-scale IGM structure predicted by hydrodynamic simulations (\cite{Mesinger15, Choudhury15}).  This approach has been used to study the impact of IGM damping wing absorption to Lyman alpha emitting galaxies, which can be sensitive to neutral hydrogen structure over a wide range of scales.  As another example, \cite{Mutch16} incorporate a semi-analytic star-formation model, based on $N$-body merger trees, within a semi-numeric treatment of reionization, allowing for a more physical parametrization of source properties.

The diversification of these tools was driven by the challenge of finding the right tool for the job.  There is no ``one size fits all'' for EoR simulations.  Each analysis should be tailored to a given observable, focusing on the most relevant physics and scales (e.g. small-scale IGM structure, Ly$\alpha$ forest, large-scale EoR structure, ISM sub-structure, molecular cooled halos, etc.), and parametrizing the missing physics.  In the next section, we will discuss how applying these simulation tools to current observations allowed us to start narrowing down the history of reionization.

\section{The current state of knowledge: timing of reionization}
\label{sec:timing}

Our current knowledge about the EoR stems from two classes of probes: (i) integral constraints from the CMB; and (ii) astrophysical ``flashlights'' which illuminate the intervening IGM.  The most important CMB constraint comes in the form of the Thompson scattering optical depth to the last scattering surface, $\tau_e$.\footnote{Alternative probes such as E-mode polarization as a function of angular scale (e.g. \cite{MH08}), the patchiness of $\tau_e$ (e.g. \cite{DS09}), the kinetic Sunyaev-Zel'dovich signal from patchy reionization (e.g. \cite{MMS12}) , could yield interesting results in the future provided systematics can be controlled (see the review of \cite{Reichardt16}).}  The latest results from the {\it Planck} satellite give $\tau_e = 0.058 \pm 0.012 ~ (1\sigma)$ \cite{Planck16}, implying a later EoR than previous estimates.  Since $\tau_e$ is an integral constraint, using it to infer the EoR requires the assumption of a functional form for the redshift evolution of the average neutral fraction, $\avenf$(z).

On the other hand, if found at a sufficiently high-$z$, astrophysical flashlights allow us to glimpse the state of the EoR at a given redshift.  The ``color'' of these flashlights is generally Lyman alpha, which is the strongest emission line and is sensitive to the presence of neutral hydrogen.  While Lyman alpha absorption in the IGM produces a Ly$\alpha$ forest in the spectra of moderate redshift QSOs, the line saturates quickly beyond $z\gsim6$ (e.g. \cite{fan06}).  However, during the EoR, the Lorentzian damping wings of the line can provide a sufficiently high optical depth to absorb photons on the red side of the Lyman alpha resonance.  The imprint of this EoR damping wing can be studied in individual bright QSO spectra (e.g. \cite{Bolton11}), or in ensemble-averaged properties of fainter galaxy spectra (e.g. \cite{Dijkstra14Review}).

\begin{figure}
{
\begin{center}
  \includegraphics[width=0.8\textwidth]{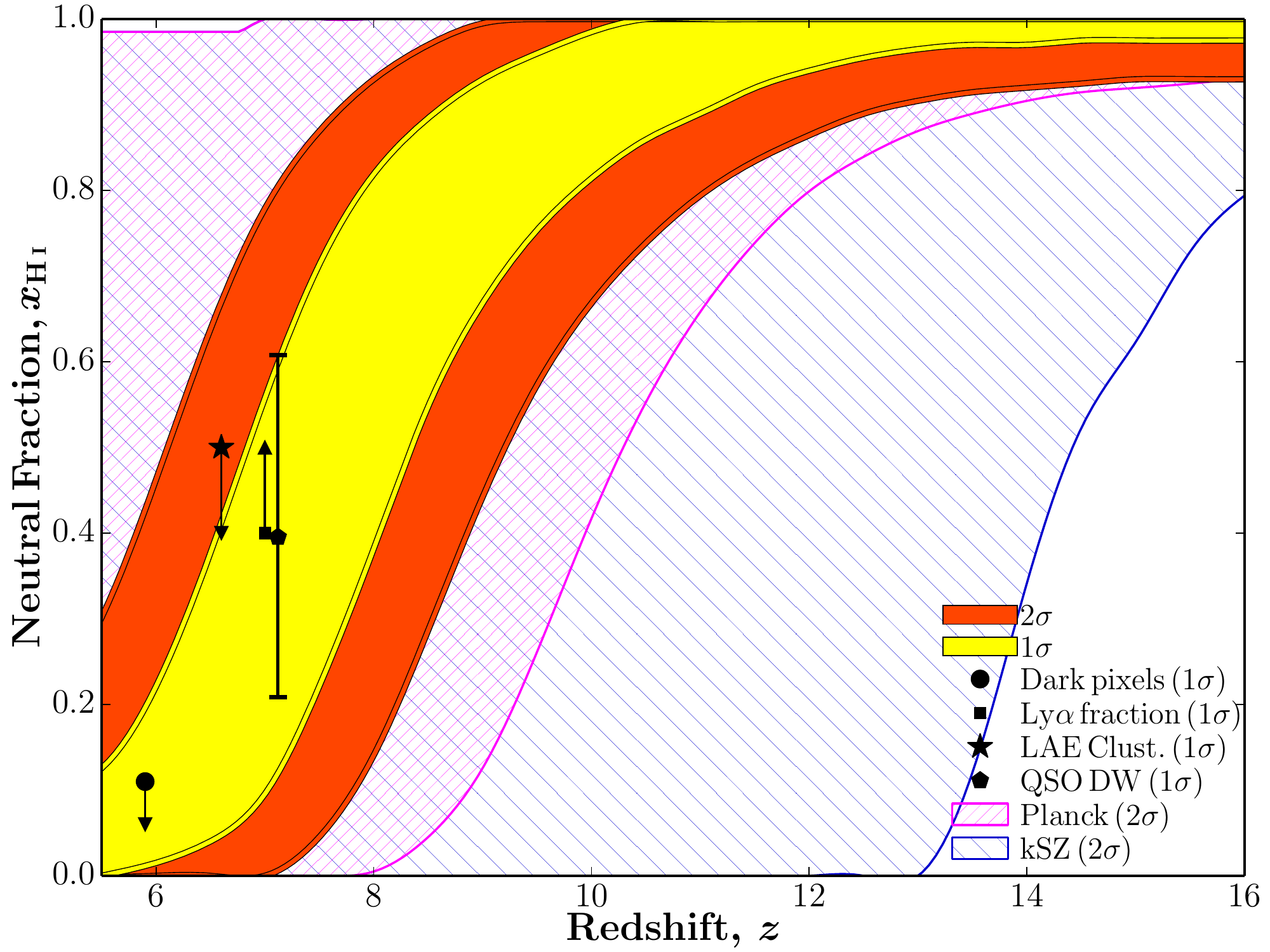}
\end{center}
}
\caption{
  The global history of reionization.  A physically-motivated EoR model was sampled, with the likelihood of each resulting $\avenf(z)$ curve provided by current observations.  The figure is taken from \cite{GM17} (see also similar results by \cite{MCF15, PTC16}).
}
\label{fig:EoRhistory}
\vspace{-0.5\baselineskip}
\end{figure}

In Fig. \ref{fig:EoRhistory}, we summarize the current state of knowledge on the history of reionization (taken from \cite{GM17}; see also similar results by \cite{MCF15, PTC16}).
Fitting a physically-motivated basis set of $\avenf(z)$ to current observations, these authors constrain the epochs corresponding to an average neutral fraction of (75, 50, 25) per cent, to $z= (8.52\substack{+0.96 \\ -0.87}, 7.57\substack{+0.78 \\ -0.73}, 6.82\substack{+0.78 \\ -0.71})$, (1-$\sigma$).  The strongest constraints here come from the {\it first detection of ongoing reionization}, obtained from the spectra of the $z=7.1$ QSOs ULASJ1120+0641: $\avenf(z=7.1) = 0.4^{+0.41}_{-0.32}$ (2-$\sigma$); see also the recent work by \cite{Mason17} who obtain comparable limits from the disappearance of Lyman alpha emitting galaxies beyond $z\gsim6$ (not shown in the figure).

The past few years have witnessed dramatic progress in our understanding of the timing of reionization.  However, virtually nothing is known about the sources (and sinks) which govern this process.  The true revolution in our understanding of the first billion years will come with upcoming 21-cm interferometers.

\section{Future potential: the cosmic 21cm signal}
\label{sec:21cm}

The spin flip transition of HI, resulting in the emission of a photon with a wavelength of 21-cm, is an extremely powerful probe of the EoR and CD (see the review of \cite{FOB06}).  The cosmic neutral hydrogen can be seen in contrast against the CMB.
%
%
The 21-cm signal is sensitive to both the ionization and thermal state of the cosmic gas, and it contains both astrophysical and cosmological terms.
Since it is a line signal, with a given frequency corresponding to a redshift, upcoming interferometers can provide a 3D map of the first billion years of our Universe!

Current interferometers are already taking data, hoping for a statistical detection of the EoR.  These include the Low Frequency Array (LOFAR; \cite{vanHaarlem13}),  Murchison Wide Field Array (MWA; \cite{Tingay13}), and the Precision Array for Probing the Epoch of Reionization (PAPER; \cite{Parsons10}).  Next-generation interferometers, the Hydrogen Epoch of Reionization Arrays (HERA; \cite{DeBoer16}), and the Square Kilometre Array (SKA; e.g. \cite{Koopmans15}) will be completed soon, with sufficient sensitivity to capture even the earlier stages of the CD, and could even provide tomographic maps of the first billion years. 

{\it How do we tap into this wealth of information?}
The {\it patterns} of the 21-cm signal (i.e. the fluctuations in the ionization and temperature of the IGM) encode the UV and X-ray properties of the first galaxies!  For example, if star-formation is only efficient in relatively massive galaxies, the cosmic HII regions should be relatively large and isolated, at a given stage in the EoR (e.g. \cite{FZH04, McQuinn07, Iliev12}).  Efficient self-shielding of gas clumps in the IGM can produce an opposite effect, sapping the growth of large HII regions (e.g. \cite{McQuinn07, SM14}).  Similarly, the smoothness of the earlier epoch of heating would allow us to discriminate between different high-energy processes in the first galaxies (e.g. \cite{Pacucci14, FBV14}).

\begin{figure}
{
\includegraphics[width=\textwidth]{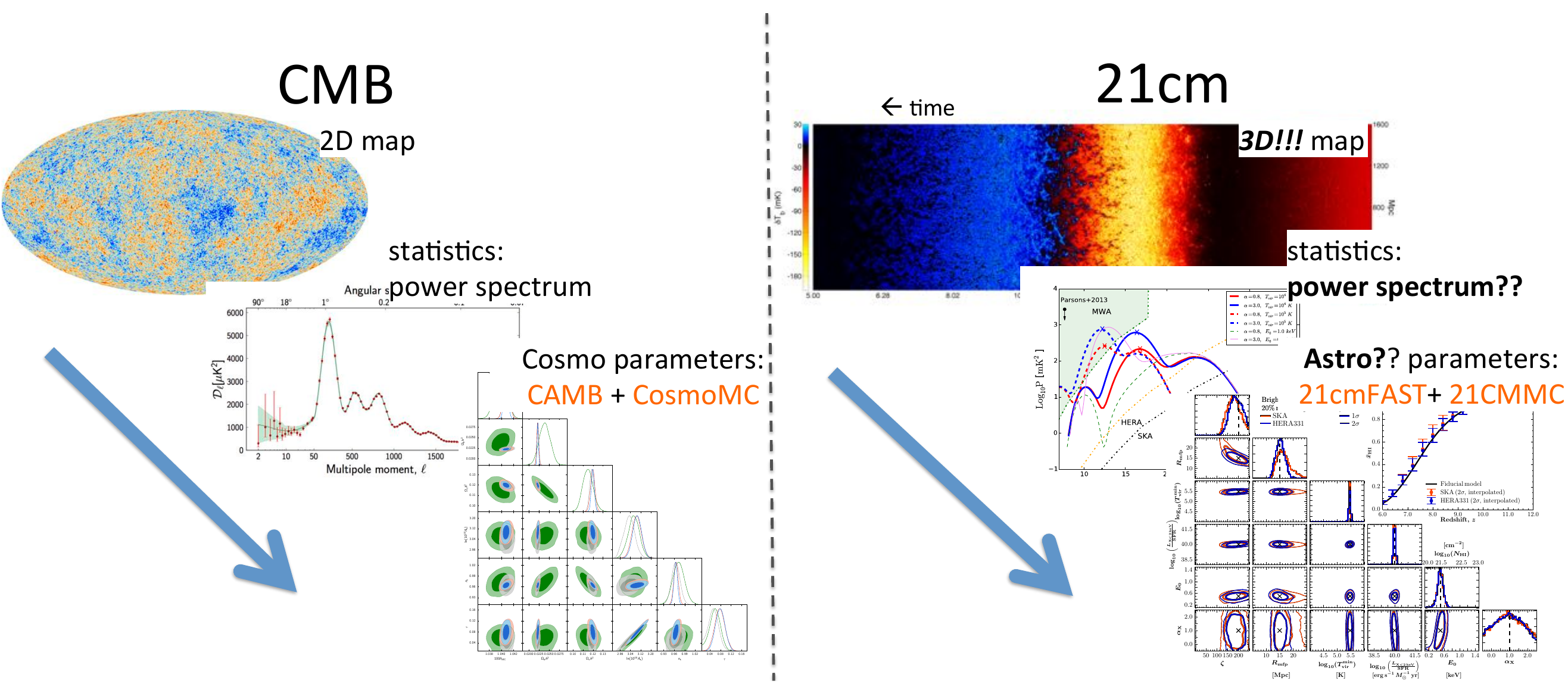}
}
\caption{
  Schematic illustrating the parallels between the established Bayesian parameter recovery with the CMB, and the planed parameter recovery with the cosmic 21cm signal.
}
\label{fig:21cmmc}
\vspace{-0.5\baselineskip}
\end{figure}

{\it How do we quantify this?}  Here we can take inspiration from the widely-successful Bayesian framework for cosmological parameter estimation from the CMB (see the left panel of Fig. \ref{fig:21cmmc}).  A 2D map of the CMB is compressed into a summary statistic (the angular power spectra).  Then an MCMC sampler (e.g. \textsc{CosmoMC}; \cite{LB02}) is used to constrain physical cosmological parameters, comparing corresponding theoretical power spectra (with, e.g. \textsc{CMBfast}/\textsc{CAMB}; \cite{SZ96, LCL00}) against the observations.

Astrophysics with the 21-cm signal is similar in many ways (see the right panel of Fig. \ref{fig:21cmmc}).  Here we start with a {\it 3D} map of the 21-cm signal, containing many orders of magnitude more modes than the CMB (for a review of the associated observational challenges, see the chapter by S. Zaroubi in this volume).  This signal is then compressed into summary statistics.  Although the power spectrum is a natural choice for a summary statistic, it is not clear whether this is optimal because the 21-cm signal (unlike the CMB) is highly non-Gaussian.  Complementary information could be provided by alternate statistics, such as the 1-point distribution (e.g. \cite{BL08, WP14}), the bispectrum (e.g. \cite{BP05, Shimabukuro16, Majumdar17}), HII region size characterizations (e.g. \cite{Zahn07, MF07, Giri18}), etc.
Finally, an MCMC sampler (e.g. \textsc{21cmMC}; \cite{GM15}) can be used to constrain {\it astrophysical} parameters, by comparing the corresponding theoretical summary statistic (e.g. using \textsc{21cmFAST}; \cite{MFC11}) against the observations.

\section{Conclusions}

The past few years have witnessed substantial progress in the modeling of the epoch of reionization and the cosmic dawn.  Simulations have not only become more sophisticated, but have evolved and diversified to best suit their intended use. We now have many techniques for radiative transfer and how the radiation is coupled to the underlying source and sink fields.

Careful application of these tools to current observations allowed us to get an idea of when reionization occurred (e.g. the midpoint is constrained to $z = 7.6_{-0.7}^{+0.8}$ at 1 $\sigma$).  The strongest constraints come from the optical depth to the CMB, and the first detection of ongoing reionization from the spectra of the $z=7.1$ QSOs ULASJ1120+0641.

Unfortunately, our knowledge does not extend far beyond the timing.  We know virtually nothing about the galaxies, AGN, and gas clumps which are the dominant actors in the EoR and CD.  Luckily, upcoming 21-cm interferometers will open-up an astrophysical treasure trove, providing a 3D map of the process.  The properties of astrophysical sources and sinks are encoded in the patterns of the 21-cm signal. 
Bayesian frameworks are being developed which can tap into this physical bounty, providing astrophysical parameter constraints.  The era of precision {\it astrophysical cosmology} is on our doorsteps!



\begin{thebibliography}{99}

\bibitem[{Baek} et~al., 2010]{Baek10}
{Baek}, S., {Semelin}, B., {Di Matteo}, P., {Revaz}, Y., and {Combes}, F.
  (2010).
\newblock {Reionization by UV or X-ray sources}.
\newblock {\em \aap}, 523:A4.

\bibitem[{Barkana} and {Loeb}, 2008]{BL08}
{Barkana}, R. and {Loeb}, A. (2008).
\newblock {The difference PDF of 21-cm fluctuations: a powerful statistical
  tool for probing cosmic reionization}.
\newblock {\em \mnras}, 384:1069--1079.

\bibitem[{Bharadwaj} and {Pandey}, 2005]{BP05}
{Bharadwaj}, S. and {Pandey}, S.~K. (2005).
\newblock {Probing non-Gaussian features in the HI distribution at the epoch of
  re-ionization}.
\newblock {\em \mnras}, 358:968--976.

\bibitem[{Bolton} et~al., 2011]{Bolton11}
{Bolton}, J.~S., {Haehnelt}, M.~G., {Warren}, S.~J., {Hewett}, P.~C.,
  {Mortlock}, D.~J., {Venemans}, B.~P., {McMahon}, R.~G., and {Simpson}, C.
  (2011).
\newblock {How neutral is the intergalactic medium surrounding the redshift z =
  7.085 quasar ULAS J1120+0641?}
\newblock {\em \mnras}, 416:L70--L74.

\bibitem[{Bond} et~al., 1991]{Bond91}
{Bond}, J.~R., {Cole}, S., {Efstathiou}, G., and {Kaiser}, N. (1991).
\newblock {Excursion set mass functions for hierarchical Gaussian
  fluctuations}.
\newblock {\em \apj}, 379:440--460.

\bibitem[{Choudhury} et~al., 2009]{CHR09}
{Choudhury}, T.~R., {Haehnelt}, M.~G., and {Regan}, J. (2009).
\newblock {Inside-out or outside-in: the topology of reionization in the
  photon-starved regime suggested by Ly{$\alpha$} forest data}.
\newblock {\em \mnras}, 394:960--977.

\bibitem[{Choudhury} et~al., 2015]{Choudhury15}
{Choudhury}, T.~R., {Puchwein}, E., {Haehnelt}, M.~G., and {Bolton}, J.~S.
  (2015).
\newblock {Lyman {$\alpha$} emitters gone missing: evidence for late
  reionization?}
\newblock {\em \mnras}, 452:261--277.

\bibitem[{DeBoer} et~al., 2016]{DeBoer16}
{DeBoer}, D.~R. et~al. (2016).
\newblock {Hydrogen Epoch of Reionization Array (HERA)}.
\newblock {\em ArXiv e-prints:1606.07473}.

\bibitem[{Dijkstra}, 2014]{Dijkstra14Review}
{Dijkstra}, M. (2014).
\newblock {Lyman Alpha Emitting Galaxies as a Probe of Reionization}.
\newblock {\em ArXiv e-prints:1406.7292}.

\bibitem[{Dixon} et~al., 2016]{Dixon15}
{Dixon}, K.~L., {Iliev}, I.~T., {Mellema}, G., {Ahn}, K., and {Shapiro}, P.~R.
  (2016).
\newblock {The large-scale observational signatures of low-mass galaxies during
  reionization}.
\newblock {\em \mnras}, 456:3011--3029.

\bibitem[{Dvorkin} and {Smith}, 2009]{DS09}
{Dvorkin}, C. and {Smith}, K.~M. (2009).
\newblock {Reconstructing patchy reionization from the cosmic microwave
  background}.
\newblock {\em \prd}, 79(4):043003--+.

\bibitem[Fan et~al., 2006]{fan06}
Fan, X. et~al. (2006).
\newblock {Constraining the Evolution of the Ionizing Background and the Epoch
  of Reionization with z\~{}6 Quasars. II. A Sample of 19 Quasars}.
\newblock {\em \aj}, 132:117--136.

\bibitem[{Fialkov} et~al., 2014]{FBV14}
{Fialkov}, A., {Barkana}, R., and {Visbal}, E. (2014).
\newblock {The observable signature of late heating of the Universe during
  cosmic reionization}.
\newblock {\em \nat}, 506:197--199.

\bibitem[{Fialkov} et~al., 2013]{Fialkov13}
{Fialkov}, A., {Barkana}, R., {Visbal}, E., {Tseliakhovich}, D., and {Hirata},
  C.~M. (2013).
\newblock {The 21-cm signature of the first stars during the Lyman-Werner
  feedback era}.
\newblock {\em \mnras}, 432:2909--2916.

\bibitem[{Furlanetto} et~al., 2004a]{FHZ04}
{Furlanetto}, S.~R., {Hernquist}, L., and {Zaldarriaga}, M. (2004a).
\newblock {Constraining the topology of reionization through Ly{$\alpha$}
  absorption}.
\newblock {\em \mnras}, 354:695--707.

\bibitem[{Furlanetto} et~al., 2006]{FOB06}
{Furlanetto}, S.~R., {Oh}, S.~P., and {Briggs}, F.~H. (2006).
\newblock {Cosmology at low frequencies: The 21 cm transition and the
  high-redshift Universe}.
\newblock {\em \physrep}, 433:181--301.

\bibitem[{Furlanetto} et~al., 2004b]{FZH04}
{Furlanetto}, S.~R., {Zaldarriaga}, M., and {Hernquist}, L. (2004b).
\newblock {The Growth of H II Regions During Reionization}.
\newblock {\em \apj}, 613:1--15.

\bibitem[{Geil} and {Wyithe}, 2008]{GW08}
{Geil}, P.~M. and {Wyithe}, J.~S.~B. (2008).
\newblock {The impact of a percolating IGM on redshifted 21-cm observations of
  quasar HII regions}.
\newblock {\em \mnras}, 386:1683--1694.

\bibitem[{Giri} et~al., 2018]{Giri18}
{Giri}, S.~K., {Mellema}, G., {Dixon}, K.~L., and {Iliev}, I.~T. (2018).
\newblock {Bubble size statistics during reionization from 21-cm tomography}.
\newblock {\em \mnras}, 473:2949--2964.

\bibitem[{Gnedin} and {Abel}, 2001]{GA01}
{Gnedin}, N.~Y. and {Abel}, T. (2001).
\newblock {Multi-dimensional cosmological radiative transfer with a Variable
  Eddington Tensor formalism}.
\newblock {\em New Astronomy}, 6:437--455.

\bibitem[{Greig} and {Mesinger}, 2015]{GM15}
{Greig}, B. and {Mesinger}, A. (2015).
\newblock {21CMMC: an MCMC analysis tool enabling astrophysical parameter
  studies of the cosmic 21 cm signal}.
\newblock {\em \mnras}, 449:4246--4263.

\bibitem[{Greig} and {Mesinger}, 2017]{GM17}
{Greig}, B. and {Mesinger}, A. (2017).
\newblock {The global history of reionization}.
\newblock {\em \mnras}, 465:4838--4852.

\bibitem[Iliev et~al., 2006]{Iliev06}
Iliev, I.~T. et~al. (2006).
\newblock {Cosmological radiative transfer codes comparison project - I. The
  static density field tests}.
\newblock {\em \mnras}, 371:1057--1086.

\bibitem[{Iliev} et~al., 2014]{Iliev14}
{Iliev}, I.~T., {Mellema}, G., {Ahn}, K., {Shapiro}, P.~R., {Mao}, Y., and
  {Pen}, U.-L. (2014).
\newblock {Simulating cosmic reionization: how large a volume is large enough?}
\newblock {\em \mnras}, 439:725--743.

\bibitem[{Iliev} et~al., 2012]{Iliev12}
{Iliev}, I.~T., {Mellema}, G., {Shapiro}, P.~R., {Pen}, U.-L., {Mao}, Y.,
  {Koda}, J., and {Ahn}, K. (2012).
\newblock {Can 21-cm observations discriminate between high-mass and low-mass
  galaxies as reionization sources?}
\newblock {\em \mnras}, 423:2222--2253.

\bibitem[{Kakiichi} et~al., 2017]{Kakiichi17}
{Kakiichi}, K., {Graziani}, L., {Ciardi}, B., {Meiksin}, A., {Compostella}, M.,
  {Eide}, M.~B., and {Zaroubi}, S. (2017).
\newblock {The concerted impact of galaxies and QSOs on the ionization and
  thermal state of the intergalactic medium}.
\newblock {\em \mnras}, 468:3718--3736.

\bibitem[{Koopmans} et~al., 2015]{Koopmans15}
{Koopmans}, L. et~al. (2015).
\newblock {The Cosmic Dawn and Epoch of Reionisation with SKA}.
\newblock {\em Advancing Astrophysics with the Square Kilometre Array
  (AASKA14)}, page~1.

\bibitem[{Lacey} and {Cole}, 1993]{LC93}
{Lacey}, C. and {Cole}, S. (1993).
\newblock {Merger rates in hierarchical models of galaxy formation}.
\newblock {\em \mnras}, 262:627--649.

\bibitem[{Lewis} and {Bridle}, 2002]{LB02}
{Lewis}, A. and {Bridle}, S. (2002).
\newblock {Cosmological parameters from CMB and other data: A Monte Carlo
  approach}.
\newblock {\em \prd}, 66(10):103511.

\bibitem[{Lewis} et~al., 2000]{LCL00}
{Lewis}, A., {Challinor}, A., and {Lasenby}, A. (2000).
\newblock {Efficient Computation of Cosmic Microwave Background Anisotropies in
  Closed Friedmann-Robertson-Walker Models}.
\newblock {\em \apj}, 538:473--476.

\bibitem[{Majumdar} et~al., 2017]{Majumdar17}
{Majumdar}, S., {Pritchard}, J.~R., {Mondal}, R., {Watkinson}, C.~A.,
  {Bharadwaj}, S., and {Mellema}, G. (2017).
\newblock {Quantifying the non-Gaussianity in the EoR 21-cm signal through
  bispectrum}.
\newblock {\em ArXiv e-prints}.

\bibitem[{Mason} et~al., 2017]{Mason17}
{Mason}, C.~A., {Treu}, T., {Dijkstra}, M., {Mesinger}, A., {Trenti}, M.,
  {Pentericci}, L., {de Barros}, S., and {Vanzella}, E. (2017).
\newblock {The Universe is Reionizing at z\~{}7: Bayesian Inference of the IGM
  Neutral Fraction Using Ly$\alpha$ Emission from Galaxies}.
\newblock {\em ArXiv e-prints:1709.05356}.

\bibitem[{McQuinn} et~al., 2007]{McQuinn07}
{McQuinn}, M., {Lidz}, A., {Zahn}, O., {Dutta}, S., {Hernquist}, L., and
  {Zaldarriaga}, M. (2007).
\newblock {The morphology of HII regions during reionization}.
\newblock {\em \mnras}, 377:1043--1063.

\bibitem[{Mesinger}, 2010]{Mesinger10}
{Mesinger}, A. (2010).
\newblock {Was reionization complete by z \~{} 5-6?}
\newblock {\em \mnras}, 407:1328--1337.

\bibitem[{Mesinger} et~al., 2015]{Mesinger15}
{Mesinger}, A., {Aykutalp}, A., {Vanzella}, E., {Pentericci}, L., {Ferrara},
  A., and {Dijkstra}, M. (2015).
\newblock {Can the intergalactic medium cause a rapid drop in Ly{$\alpha$}
  emission at z = 6?}
\newblock {\em \mnras}, 446:566--577.

\bibitem[{Mesinger} and {Furlanetto}, 2007]{MF07}
{Mesinger}, A. and {Furlanetto}, S. (2007).
\newblock {Efficient Simulations of Early Structure Formation and
  Reionization}.
\newblock {\em \apj}, 669:663--675.

\bibitem[{Mesinger} et~al., 2011]{MFC11}
{Mesinger}, A., {Furlanetto}, S., and {Cen}, R. (2011).
\newblock {21CMFAST: a fast, seminumerical simulation of the high-redshift
  21-cm signal}.
\newblock {\em \mnras}, 411:955--972.

\bibitem[{Mesinger} et~al., 2016]{MGS16}
{Mesinger}, A., {Greig}, B., and {Sobacchi}, E. (2016).
\newblock {The Evolution Of 21 cm Structure (EOS): public, large-scale
  simulations of Cosmic Dawn and reionization}.
\newblock {\em \mnras}, 459:2342--2353.

\bibitem[{Mesinger} et~al., 2012]{MMS12}
{Mesinger}, A., {McQuinn}, M., and {Spergel}, D.~N. (2012).
\newblock {The kinetic Sunyaev-Zel'dovich signal from inhomogeneous
  reionization: a parameter space study}.
\newblock {\em \mnras}, 422:1403--1417.

\bibitem[{Mitra} et~al., 2015]{MCF15}
{Mitra}, S., {Choudhury}, T.~R., and {Ferrara}, A. (2015).
\newblock {Cosmic reionization after Planck}.
\newblock {\em \mnras}, 454:L76--L80.

\bibitem[{Mortonson} and {Hu}, 2008]{MH08}
{Mortonson}, M.~J. and {Hu}, W. (2008).
\newblock {Model-Independent Constraints on Reionization from Large-Scale
  Cosmic Microwave Background Polarization}.
\newblock {\em \apj}, 672:737--751.

\bibitem[{Mutch} et~al., 2016]{Mutch16}
{Mutch}, S.~J., {Geil}, P.~M., {Poole}, G.~B., {Angel}, P.~W., {Duffy}, A.~R.,
  {Mesinger}, A., and {Wyithe}, J.~S.~B. (2016).
\newblock {Dark-ages reionization and galaxy formation simulation III:
  Modelling galaxy formation and the Epoch of Reionization}.
\newblock {\em \mnras}, 462:250--276.

\bibitem[{Ocvirk} et~al., 2016]{Ocvirk16}
{Ocvirk}, P. et~al. (2016).
\newblock {Cosmic Dawn (CoDa): the First Radiation-Hydrodynamics Simulation of
  Reionization and Galaxy Formation in the Local Universe}.
\newblock {\em \mnras}, 463:1462--1485.

\bibitem[{Pacucci} et~al., 2014]{Pacucci14}
{Pacucci}, F., {Mesinger}, A., {Mineo}, S., and {Ferrara}, A. (2014).
\newblock {The X-ray spectra of the first galaxies: 21 cm signatures}.
\newblock {\em \mnras}, 443:678--686.

\bibitem[Parsons et~al., 2010]{Parsons10}
Parsons, A.~R. et~al. (2010).
\newblock {The Precision Array for Probing the Epoch of Re-ionization: Eight
  Station Results}.
\newblock {\em \aj}, 139:1468--1480.

\bibitem[{Partl} et~al., 2011]{Partl11}
{Partl}, A.~M., {Maselli}, A., {Ciardi}, B., {Ferrara}, A., and {M{\"u}ller},
  V. (2011).
\newblock {Enabling parallel computing in CRASH}.
\newblock {\em \mnras}, 414:428--444.

\bibitem[{Planck Collaboration}, 2016]{Planck16}
{Planck Collaboration}, . (2016).
\newblock {Planck 2016 intermediate results. XLVII. Planck constraints on
  reionization history}.
\newblock {\em ArXiv e-prints:1605.03507}.

\bibitem[{Price} et~al., 2016]{PTC16}
{Price}, L.~C., {Trac}, H., and {Cen}, R. (2016).
\newblock {Reconstructing the redshift evolution of escaped ionizing flux from
  early galaxies with Planck and HST observations}.
\newblock {\em ArXiv e-prints:1605.03970}.

\bibitem[{Reichardt}, 2016]{Reichardt16}
{Reichardt}, C.~L. (2016).
\newblock {Observing the Epoch of Reionization with the Cosmic Microwave
  Background}.
\newblock In {Mesinger}, A., editor, {\em Astrophysics and Space Science
  Library}, volume 423 of {\em Astrophysics and Space Science Library}, page
  227.

\bibitem[{Santos} et~al., 2010]{Santos10}
{Santos}, M.~G., {Ferramacho}, L., {Silva}, M.~B., {Amblard}, A., and {Cooray},
  A. (2010).
\newblock {Fast large volume simulations of the 21-cm signal from the
  reionization and pre-reionization epochs}.
\newblock {\em \mnras}, 406:2421--2432.

\bibitem[{Seljak} and {Zaldarriaga}, 1996]{SZ96}
{Seljak}, U. and {Zaldarriaga}, M. (1996).
\newblock {A Line-of-Sight Integration Approach to Cosmic Microwave Background
  Anisotropies}.
\newblock {\em \apj}, 469:437.

\bibitem[{Shimabukuro} et~al., 2016]{Shimabukuro16}
{Shimabukuro}, H., {Yoshiura}, S., {Takahashi}, K., {Yokoyama}, S., and
  {Ichiki}, K. (2016).
\newblock {21 cm line bispectrum as a method to probe cosmic dawn and epoch of
  reionization}.
\newblock {\em \mnras}, 458:3003--3011.

\bibitem[{Sobacchi} and {Mesinger}, 2014]{SM14}
{Sobacchi}, E. and {Mesinger}, A. (2014).
\newblock {Inhomogeneous recombinations during cosmic reionization}.
\newblock {\em \mnras}, 440:1662--1673.

\bibitem[{Tingay} et~al., 2013]{Tingay13}
{Tingay}, S.~J. et~al. (2013).
\newblock {The Murchison Widefield Array: The Square Kilometre Array Precursor
  at Low Radio Frequencies}.
\newblock {\em \pasa}, 30:7.

\bibitem[{Trac} and {Cen}, 2007]{TC07}
{Trac}, H. and {Cen}, R. (2007).
\newblock {Radiative Transfer Simulations of Cosmic Reionization. I.
  Methodology and Initial Results}.
\newblock {\em \apj}, 671:1--13.

\bibitem[{Trac} and {Gnedin}, 2011]{TG11}
{Trac}, H.~Y. and {Gnedin}, N.~Y. (2011).
\newblock {Computer Simulations of Cosmic Reionization}.
\newblock {\em Advanced Science Letters}, 4:228--243.

\bibitem[{van Haarlem} et~al., 2013]{vanHaarlem13}
{van Haarlem}, M.~P. et~al. (2013).
\newblock {LOFAR: The LOw-Frequency ARray}.
\newblock {\em \aap}, 556:A2.

\bibitem[{Watkinson} and {Pritchard}, 2014]{WP14}
{Watkinson}, C.~A. and {Pritchard}, J.~R. (2014).
\newblock {Distinguishing models of reionization using future radio
  observations of 21-cm 1-point statistics}.
\newblock {\em \mnras}, 443:3090--3106.

\bibitem[{Xu} et~al., 2016]{Xu16}
{Xu}, H., {Wise}, J.~H., {Norman}, M.~L., {Ahn}, K., and {O'Shea}, B.~W.
  (2016).
\newblock {Galaxy Properties and UV Escape Fractions during the Epoch of
  Reionization: Results from the Renaissance Simulations}.
\newblock {\em \apj}, 833:84.

\bibitem[{Zahn} et~al., 2007]{Zahn07}
{Zahn}, O., {Lidz}, A., {McQuinn}, M., {Dutta}, S., {Hernquist}, L.,
  {Zaldarriaga}, M., and {Furlanetto}, S.~R. (2007).
\newblock {Simulations and Analytic Calculations of Bubble Growth during
  Hydrogen Reionization}.
\newblock {\em \apj}, 654:12--26.

\bibitem[{Zahn} et~al., 2011]{Zahn11}
{Zahn}, O., {Mesinger}, A., {McQuinn}, M., {Trac}, H., {Cen}, R., and
  {Hernquist}, L.~E. (2011).
\newblock {Comparison of reionization models: radiative transfer simulations
  and approximate, seminumeric models}.
\newblock {\em \mnras}, 414:727--738.

\end{thebibliography}
\end{document}